\documentclass[prd,twocolumn,nopacs,floatfix,amsmath,nofootinbib,amssymb,floatfix]{revtex4}
\usepackage{graphicx,color,dcolumn,booktabs,bm}
\usepackage{longtable,lscape}
\usepackage{pdfpages}
\usepackage{txfonts}
\usepackage{subfigure}
\usepackage{overpic}
\usepackage{amssymb}
\usepackage{indentfirst}
\usepackage{feynmf}   
\usepackage{slashed}  
\usepackage{cases}
\usepackage{color}
\usepackage{multirow}
\usepackage{threeparttable}
\usepackage{epstopdf}
\usepackage{enumerate}
\usepackage{graphicx,color,dcolumn,booktabs,bm}
\usepackage[colorlinks, citecolor=blue,anchorcolor=red,menucolor=red, linkcolor=red,filecolor=red,urlcolor=blue,frenchlinks=red]{hyperref}

\graphicspath{{Figures/}} %

\begin{document}

\title{Universal behavior of mass gaps existing in the single heavy baryon family}
\author{Bing Chen$^{1,3}$}\email{chenbing@ahstu.edu.cn}
\author{Si-Qiang Luo$^{2,3,4}$}\email{luosq15@lzu.edu.cn}
\author{Xiang Liu$^{2,3,4}$\footnote{Corresponding author}}\email{xiangliu@lzu.edu.cn}
\affiliation{$^1$School of Electrical and Electronic Engineering, Anhui Science and Technology University, Bengbu 233000, China\\
$^2$School of Physical Science and Technology, Lanzhou University, Lanzhou 730000, China\\
$^3$Lanzhou Center for Theoretical Physics, Key Laboratory of Theoretical Physics of Gansu Province, and Frontiers Science Center for Rare Isotopes, Lanzhou University, Lanzhou 730000, China\\
$^4$Research Center for Hadron and CSR Physics, Lanzhou University $\&$ Institute of Modern Physics of CAS, Lanzhou 730000, China}

\date{\today}

\begin{abstract}
The mass gaps existing in the discovered single heavy flavor baryons are analyzed, which show
some universal behaviors. Under the framework of a constituent quark model, we quantitatively explain why such interesting phenomenon happens, when these established excited heavy baryons are regarded as the $\lambda$-mode excitations.
Based on the universal behaviors of the discussed mass gaps, we may have three implications including the prediction of the masses of excited $\Xi_b^0$ baryons which are still missing in the experiment. For completeness, we also discuss the mass gaps of these $\rho$-mode excited single heavy flavor baryons.
\end{abstract}
\maketitle

\section{Introduction}\label{sec1}

As an effective approach to solve the nonperturbative problem of strong interaction,
studying hadron spectroscopy has become an active research issue with the abundant observations of new hadronic states in experiment (see Refs.~\cite{Chen:2016qju,Ali:2017jda,Esposito:2016noz,Lebed:2016hpi,Olsen:2017bmm,Guo:2017jvc,Liu:2019zoy,Brambilla:2019esw} for the recent progress).
Among different research aspects of the study of hadron spectroscopy, mass spectrum analysis is a crucial  way to decode the property of hadronic states.

Until now, there have been different methods to perform mass spectrum analysis. If seriously depicting the mass spectrum, we have various versions of potential model~\cite{Eichten:1978tg,Godfrey:1985xj,Capstick:1986bm,Vijande:2004he,Ebert:2009ub,Ebert:2009ua,Ebert:2011jc,Ebert:2011kk}, the flux tube model~\cite{Isgur:1983wj,Isgur:1984bm}, the quantum chromodynamics (QCD) sum rule~\cite{Shifman:1978bx,Shifman:1978by}, the lattice QCD~\cite{Dudek:2010wm,Dudek:2011tt,Edwards:2011jj}, and so on.
Besides, some semi-quantitative methods were extensively applied to the mass spectrum analysis. A typical example is the Regge trajectory analysis~\cite{Regge:1959mz,Regge:1960zc,Chew:1961ev,Chew:1962eu} which has been adopted to investigate the mass spectrum of different kinds of hadrons~\cite{Anisovich:2000kxa,Brisudova:2003dj,Zhang:2004cd,Li:2007px,Wei:2010zza,Guo:2008he,Wei:2015gsa,Wei:2016jyk}.
Recently, Chen gave a mass formula for light meson and baryon, and discussed its implication~\cite{Chen:2020gml}.

On other hand, some mass gap relations existing in the hadron mass spectrum have also been realized by the theorists, which can become the simple but effective approach when scaling mass spectrum. For illustrating this point, we review several representative recent work. The similar dynamics of the $\omega$ and $\phi$ meson families requires the similarity between $\omega$ and $\phi$ meson families, where the mass gap of $\omega(782)$ and $\omega(1420)$ is similar to that of $\phi(1020)$ and $\phi(1680)$. Adopted this mass gap relation for higher states, the $Y(1915)$ state as the partner of $Y(2175)$ was predicted~\cite{Wang:2012wa}. In Ref.~\cite{He:2014xna}, Lanzhou group predicted the existence of a narrow charmonium $\psi(4S)$ with the mass around 4.263 GeV, which was estimated by the mass gap between $\Upsilon(3S)$ and $\Upsilon(4S)$, if considering the mass gap relation $m_{\psi(4S)}-m_{\psi(3S)}=m_{\Upsilon(4S)}-m_{\Upsilon(3S)}$.
In Ref.~\cite{Arifi:2020yfp}, the authors pointed out that the excitation energies of the first excited $J^P=1/2^+$ baryons with various flavor contents are about 500 MeV. As the last example, there exists a mass relation $M_{0^+}+3M_{1^+}+5M_{2^+}\simeq9M_{1^{+\prime}}$~\cite{Burns:2011fu} for charmonium and bottomonium families, where the spin-parity quantum numbers are given to distinguish the different charmonium masses with the corresponding quantum number. Accordingly, Chang {\it et al.} suggested a new mass relation $M_{0^+}+5M_{2^+}=3(M_{1^{+\prime}}+M_{1^+})$ for the $P$-wave $B_c$ mesons~\cite{Chang:2021esm}.

\begin{figure*}[htbp]
\begin{center}
\includegraphics[width=17cm,keepaspectratio]{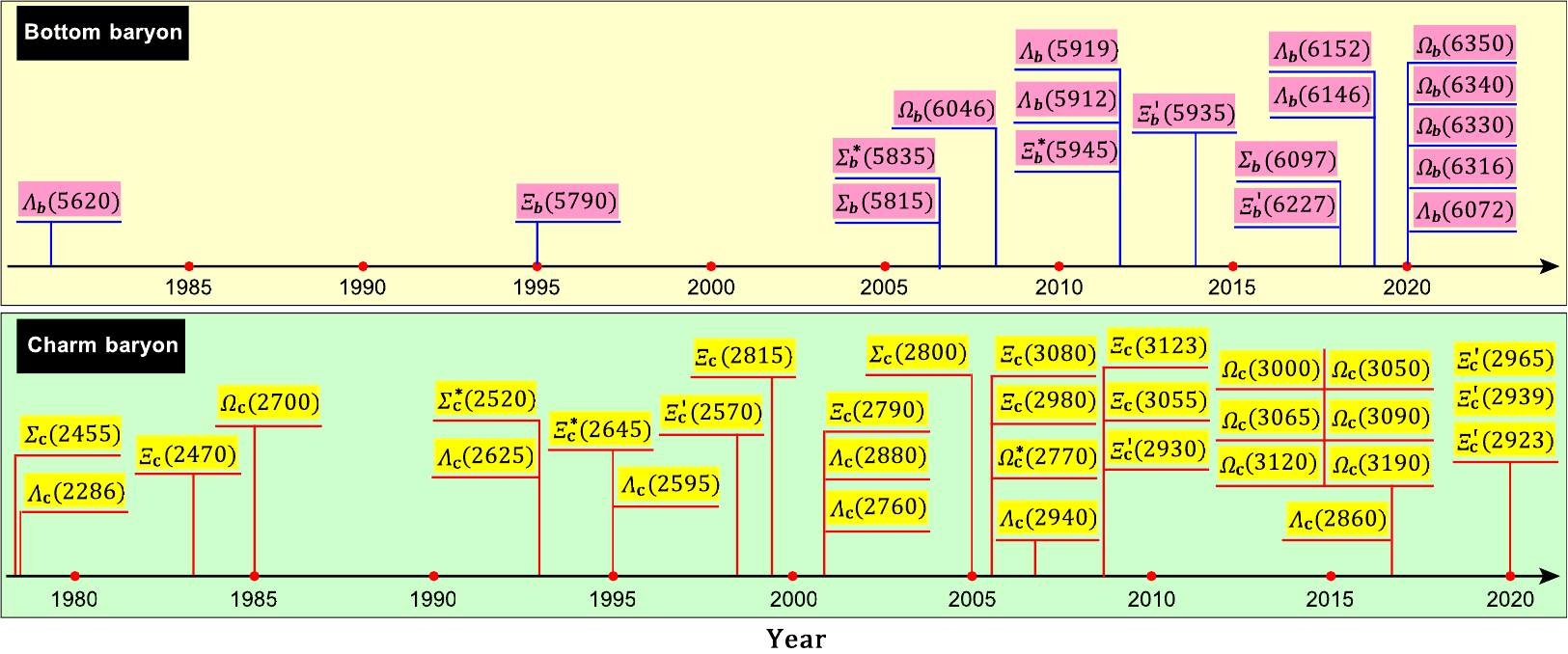}
\caption{All observed charm and bottom baryons~\cite{Aaij:2020yyt,Aaij:2020cex,Aaij:2020rkw,Zyla:2020zbs}.}\label{Fig1}
\end{center}
\end{figure*}

In the following, we need to pay attention to the single heavy flavor baryon.
Among this fantastic hadron zoo, the single heavy flavor baryon family is being constructed step by step, which is due to the big progress on the observations of charm and bottom baryons, where the LHCb Collaboration has played an important role \cite{Aaij:2012da,Aaij:2014yka,Aaij:2017vbw,Aaij:2017nav,Aaij:2018yqz,Aaij:2018tnn,Aaij:2019amv,Aaij:2020yyt,Aaij:2020rkw,Aaij:2020cex,Aaij:2020fxj}  in the past years. In Fig. \ref{Fig1}, we list these reported single heavy flavor baryon states.
Obviously, the present situation of single heavy flavor baryons shows that it is a good opportunity to check
whether there exist some mass gap relations for single heavy flavor baryon family, which will be one of tasks of this work.
We will analyze the mass gaps which are extracted by these measured mass spectrum of single heavy flavor baryons, by which we may find some universal behavior of mass gaps for these discovered single heavy baryons.

Facing such interesting mass gap phenomenon, we want to further study why there exists this universal behavior of mass gaps, which is involved in the dynamics of single heavy flavor baryon.
In this work, we start with a constituent quark model~\cite{Chen:2016iyi}, which has been successfully applied to depict the mass spectrum of single heavy baryon in our previous works~\cite{Chen:2017gnu,Chen:2018orb,Chen:2018vuc,Chen:2019ywy}. Here, two light quarks in single heavy flavor baryon are treated as a cluster, and then the single heavy flavor baryon is simplified as a quasi-two-body system.

For understanding the universal behavior of mass gaps, we perform a study of
mass gaps of single heavy baryon under the framework of constituent quark model, which shows that this universal behavior of mass gaps can be well explained. For phenomenological study, mass gap relation is always a welcome approach. Thus, in this work, we continue to apply this universal behavior of mass gaps to predict some higher single heavy baryon states, which can be tested by the future experiments.

The paper is organized as follows. After Introduction, we illustrate these mass gap relations by the measured mass spectrum of single heavy flavor baryon in Sec. \ref{sec2}. And we deduce these mass gap relations by a constituent quark model. In Sec. \ref{sec3}, its application to predict some higher states in the single heavy flavor baryon family will be given. Further discussions for the mass gaps of $\rho$-mode excited single heavy baryons will be presented in Sec. \ref{sec4}. Finally, the paper ends with a summary.

\section{Universal behavior of mass gaps}\label{sec2}
\subsection{Mass gaps of single heavy baryon}\label{sss}

The single heavy baryon system, which contains one heavy quark ($c$ or $b$ quark) and two light quarks ($u$, $d$, or $s$ quark), occupies a particular place in the whole hadron family~\cite{Roberts:2007ni}. The heavy quark in a single heavy flavor baryon system provides a quasi-static colour field for two surrounding light quarks~\cite{Korner:1994nh}. 
Based on the flavor SU(3) symmetry of light quark cluster, i.e. $3_f\otimes3_f = \bar{3}_f\oplus 6_f$, the $\Lambda_Q$ and $\Xi_Q$ baryon states belong to the $\bar{3}_f$ multiplet, while the $\Sigma_Q$, $\Xi_Q^\prime$, and $\Omega_Q$ baryon states can be grouped into the $6_f$ multiplet. Thus, in the following, we will give the mass gasps of single heavy flavor baryons in $\bar{3}_f$ and $6_f$ multiplets, and show their universal behaviors.

\subsubsection{The $\bar 3_f$ multiplet}

With the joint effort of experimentalists and theorists, the family of single heavy baryons has been established step by step. In 2017, a new charm baryon state $\Lambda_c(2860)^+$ was discovered by the LHCb Collaboration in the process $\Lambda_b^0\to\Lambda_c(2860)^+\pi^-\to{D^0p}\pi^-$ ~\cite{Aaij:2017vbw}. Combining with the previously observed $\Lambda_c(2286)^+$, $\Lambda_c(2595)^+$, $\Lambda_c(2625)^+$, $\Lambda_c(2880)^+$, $\Xi_c(2470)$, $\Xi_c(2790)$, $\Xi_c(2815)$, $\Xi_c(3055)$, and $\Xi_c(3080)$ states~\cite{Zyla:2020zbs}, all 1$S$, 2$S$, 1$P$, 1$D$ states of $\Lambda_c^+$ and $\Xi_c^{0,+}$ baryons have been discovered. In the past years, experiment has also made a big progress on searching for the excited $\Lambda_b^0$ states. In 2012, two narrow $P$-wave $\Lambda_b^0$ states, i.e., the $\Lambda_b(5912)^0$ and $\Lambda_b(5920)^0$, were first reported by the LHCb Collaboration in the $\Lambda_b^0\pi^+\pi^-$ invariant mass spectrum~\cite{Aaij:2012da}, which were confirmed by the further measurements from the CDF, CMS, and LHCb collaborations~\cite{Aaltonen:2013tta,Sirunyan:2020gtz,Aaij:2020rkw}. Besides, the $D$-wave states $\Lambda_b(6146)^0$ and $\Lambda_b(6152)^0$ have also been constructed in the recent years~\cite{Sirunyan:2020gtz,Aaij:2020rkw,Aaij:2019amv}. In last year, the $\Lambda_b(6072)^0$, which could be regarded as a good 2$S$ $\Lambda_b^0$ candidate, was found by CMS~\cite{Sirunyan:2020gtz} and LHCb~\cite{Aaij:2020rkw}. And then, the low-lying $\Lambda_c^+$, $\Xi_c^{0,+}$, and $\Lambda_b^0$ baryons including the 1$S$, 2$S$, 1$P$, and 1$D$ states have been reported by experiment. When checking the masses of these known heavy baryons (see Table \ref{table1}), the interesting universal behavior of mass gaps can be found. Specifically, the obtained mass gaps for the discussed $\Lambda_c^+$ and $\Xi_c^+$ states are nearly about 180$\sim$200 MeV (see the fourth column in Table \ref{table1}).

\begin{table}[htbp]
\caption{The measured masses of 1$S$, 2$S$, 1$P$, 1$D$ $\Lambda_c^+$ and $\Xi_c^+$ baryons~\cite{Zyla:2020zbs} and the mass gaps of corresponding states (in MeV).}\label{table1}
\renewcommand\arraystretch{1.2}
\begin{tabular*}{86mm}{c@{\extracolsep{\fill}}ccc}
\toprule[1pt]\toprule[1pt]
 $nL(J^P)$    & States                                     &  Masses                               &  Mass gap   \\
\toprule[1pt]
$1S(1/2^+)$   & $\Lambda_c(2286)^+$/$\Xi_c(2470)^+$        & 2286.5/2467.9                         & 181.4            \\
$1P(1/2^-)$   & $\Lambda_c(2595)^+$/$\Xi_c(2790)^+$        & 2592.3/2792.4                         & 200.1           \\
$1P(3/2^-)$   & $\Lambda_c(2625)^+$/$\Xi_c(2815)^+$        & 2628.1/2816.7                         & 188.6           \\
$2S(1/2^+)$   & $\Lambda_c(2765)^+$/$\Xi_c(2970)^+$        & 2766.6/2966.3                         & 199.7            \\
$1D(3/2^+)$   & $\Lambda_c(2860)^+$/$\Xi_c(3055)^+$        & 2856.1/3055.9                         & 199.8            \\
$1D(5/2^+)$   & $\Lambda_c(2880)^+$/$\Xi_c(3080)^+$        & 2881.6/3077.2                         & 195.6            \\
\bottomrule[1pt]\bottomrule[1pt]
\end{tabular*}
\end{table}

In the following, we define the following ratios
\begin{equation}
\mathcal{R}_1 = \frac{M_{2S}-M_{1S}}{\bar{M}_{1P}-M_{1S}},~~~~~~~\mathcal{R}_2 = \frac{\bar{M}_{1D}-M_{1S}}{\bar{M}_{1P}-M_{1S}}, \label{eq1}
\end{equation}
where $M_{1S}$ and $M_{2S}$ refer to the masses of the $S$-wave ground state and its first radial excitation, respectively, while $\bar{M}_{1P}$ and $\bar{M}_{1D}$ denote the spin average masses of 1$P$ and 1$D$ states. As shown in Table \ref{table2},
the values of $\mathcal{R}_1$ and $\mathcal{R}_2$ are nearly universal for the $\Lambda_c^+$ and $\Xi_c^{+}$ baryon systems. If further checking the $\Lambda_b$ baryons, the obtained $\mathcal{R}_1$ and $\mathcal{R}_2$ are also close to the corresponding values of the $\Lambda_c$ and $\Xi_c$ baryons,\footnote{We take the masses of the newly observed $\Lambda_b(6072)^0$, $\Lambda_b(6146)^0$, and $\Lambda_b(6152)^0$ states as input, where $\Lambda_b(6072)^0$, $\Lambda_b(6146)^0$, and $\Lambda_b(6152)^0$ are treated as the 2$S$ and 1$D$ states, respectively. Other assignments are also allowed for the broad resonance $\Lambda_b(6072)^0$~\cite{Xiao:2020gjo,Liang:2020kvn}.} which means the universal behavior of $\mathcal{R}_1$ and $\mathcal{R}_2$. To some extent, this novel phenomenon reflects the similar dynamics of charm and bottom baryons.

\begin{table}[t]
\caption{The extracted ratios of $\mathcal{R}_1$ and  $\mathcal{R}_2$ (see Eq.~(\ref{eq1})) for the $\Lambda_Q$ and $\Xi_Q$ baryons. The 2$S$ candidate of $\Lambda_b$ baryon with the mass around 6.07 GeV, which was found recently by the CMS~\cite{Sirunyan:2020gtz} and LHCb~\cite{Aaij:2020rkw} collaborations in the $\Lambda_b^0\pi^+\pi^-$ mass spectrum, is taken as an input.}\label{table2}
\renewcommand\arraystretch{1.2}
\begin{tabular*}{85mm}{c@{\extracolsep{\fill}}cccc}
\toprule[1pt]\toprule[1pt]
 Ratios           & $\Lambda_c^+$     & $\Xi_c$        & $\Lambda_b^0$    & $\Xi_b$      \\
\toprule[1pt]
$\mathcal{R}_1$   & 1.456             & 1.472          & 1.520          & $\cdots$            \\
$\mathcal{R}_2$   & 1.774             & 1.763          & 1.780            & $\cdots$            \\
\bottomrule[1pt]\bottomrule[1pt]
\end{tabular*}
\end{table}

\subsubsection{The $6_f$ multiplet}

In Table \ref{table3}, we extract the mass gaps according to the data of
the measured $\Sigma_Q$, $\Xi_Q^\prime$, and $\Omega_Q$ baryons.
Before discussing the mass gaps, we should briefly introduce the experimental progress on the $\Sigma_Q$, $\Xi_Q^\prime$, and $\Omega_Q$ baryons.
In the past years, some $P$-wave $\Sigma_Q$, $\Xi_Q^\prime$, and $\Omega_Q$ candidates, including the $\Xi_c(2923)^0$, $\Xi_c(2939)^0$, $\Xi_c(2965)^0$, $\Omega_c(3000)^0$, $\Omega_c(3050)^0$, $\Omega_c(3065)^0$, $\Omega_c(3090)^0$, $\Sigma_b(6097)^0$, $\Xi^\prime_b(6227)^0$, $\Omega_b(6316)^-$, $\Omega_b(6330)^-$, $\Omega_b(6340)^-$, and $\Omega_b(6350)^-$, have also been announced by the different  experiments~\cite{Li:2017uvv,Aaij:2020yyt,Aaij:2017nav,Yelton:2017qxg,Aaij:2018tnn,Aaij:2018yqz,Aaij:2020fxj,Aaij:2020cex}. These observed charm baryons can be categorized into the $6_f$ representation.
Among these observed low-lying $6_f$ heavy baryon states, the $\Xi_c(2939)^0$, $\Omega_c(3065)^0$, $\Sigma_b(6097)^0$, $\Xi^\prime_b(6227)^0$, and $\Omega_b(6350)^-$ are suggested to be the $J^P=3/2^-$ or $5/2^-$ states \cite{Cheng:2015naa,Cheng:2017ove,Chen:2017gnu,Thakkar:2016dna,Chen:2017fcs,Zhao:2017fov,Ye:2017yvl,Chen:2018orb,Chen:2018vuc,Wang:2018fjm,Aliev:2018lcs,Aliev:2018vye,Xiao:2020gjo,Lu:2020ivo,Wang:2020pri}. The $\Sigma_c(2800)^{++}$ state, which was discovered previously by the Belle Collaboration in the $e^+e^-$ collision~\cite{Mizuk:2004yu}, could also be regarded as a $P$-wave $3/2^-$ or $5/2^-$ state~\cite{Chen:2016iyi,Wang:2017kfr}.
Under these assignments, we obtain the mass gaps relevant to $\Sigma_c(2800)^{++}$, $\Xi_c(2939)^0$, and $\Omega_c(3065)^0$, which are about 130 MeV (see Table \ref{table3} for more details). Similar
value for the mass gaps of the $\Sigma_b(6097)^-$, $\Xi^\prime_b(6227)^-$, and $\Omega_b(6350)^-$ states can be found. Furthermore, the mass gaps of ground $\Sigma_Q$, $\Xi_Q^\prime$, and $\Omega_Q$ baryons are about 120 MeV.
So the last column in Table \ref{table3} also reflects the universal behavior of mass gaps of low-lying single heavy flavor baryons in $6_f$ multiplet.

\begin{table}[htbp]
\caption{The observed $\Sigma_Q$, $\Xi_Q^\prime$, and $\Omega_Q$ baryons~\cite{Zyla:2020zbs,Aaij:2020cex} and the corresponding mass gaps involved in these states (in MeV). In the last column, there are two values for each line, where the first value is the mass gap of the first and the second states listed in the second column, and the second value denotes the mass difference of the second and the third states. Here, $\Omega_b^-(\cdots)$ denotes the absent $1S$ $\Omega_b^-(3/2^+)$ state in the experiment.}\label{table3}
\renewcommand\arraystretch{1.2}
\begin{tabular*}{86mm}{l@{\extracolsep{\fill}}ll}
\toprule[1pt]\toprule[1pt]
 $nL(J^P)$    & States                                         &  Mass gap   \\
\toprule[1pt]
$1S(1/2^+)$   & $\Sigma_c(2455)^{++}/\Xi^\prime_c(2570)^+/\Omega_c(2695)^0$         & 124.4/116.8   \\
              & $\Sigma_b(5815)^+/\Xi^\prime_b(5935)^-/\Omega_b(6046)^-$            & 124.4/111.1   \\
$1S(3/2^+)$   & $\Sigma^\ast_c(2520)^{++}/\Xi^\ast_c(2645)^+/\Omega_c(2765)^0$      & 127.2/120.3    \\
              & $\Sigma^\ast_b(5835)^+/\Xi^\ast_b(5955)^-/\Omega_b^-(\cdots)$            & 125.0/~$\cdots$    \\
$1P(\frac{3}{2}^-~\textup{or}~\frac{5}{2}^-)$   & $\Sigma_c(2800)^{++}/\Xi^\prime_c(2939)^0/\Omega_c(3065)^0$         & 137.6/127.0     \\
              & $\Sigma_b(6097)^-/\Xi^\prime_b(6227)^-/\Omega_b(6350)^-$            & 128.9/123.0     \\
\bottomrule[1pt]\bottomrule[1pt]
\end{tabular*}
\end{table}

In the next section, we will explain the universal behaviors of mass gaps of single heavy flavor baryons.

\subsection{Understanding the universal behavior of mass gaps by a constituent quark model}\label{subsecA}

Although the universal behavior of mass gaps introduced in Sec. \ref{sss}  has been mentioned for many years~\cite{Chen:2014nyo,Cheng:2015naa,Cheng:2015iom}, it has never been investigated seriously. In the work, we will  give a quantitative study of the mass gap of single heavy flavor baryons with the same $nL(J^P)$ quantum numbers but different strangeness under the framework of a non-relativistic constituent quark model. Here, $n$ and $L$ denote the radial and orbital angular quantum numbers of a baryon state, respectively while $J^P$ denotes its spin-parity.

The obtained $180\sim200$ MeV mass gaps (see Table \ref{table1}) of the ground and excited states of $\Lambda_Q$ and $\Xi_Q$ baryons in $\bar 3_f$ representation reflect a universal behavior of these mass gaps. It implies that the similarity of mass gaps may be resulted from the same dynamics mechanism. Furthermore, the differences of the excited energy $E_{nL}$ for $\Lambda_Q$ and $\Xi_Q$ baryons with same $nL$ quantum number can be roughly ignored.
Such conclusion should also hold for the $6_f$ heavy baryons.
Indeed, the mass gaps shown in Sec. \ref{sss} can be naturally interpreted as the mass differences of light quark cluster involved in the discussed heavy baryons, which will be proved in following by a non-relativistic constituent quark model.

If treating two light quarks as a cluster, the single heavy baryon system can be simplified as a quasi-two-body system~\cite{Chen:2016iyi}. And then, the Hamiltonian of a heavy baryon system reads as
\begin{equation}
\hat{H} = m_Q + m_{\textup{cluster}} + \frac{\hat{p}^2}{2\mu} + V_{\textup{SI}}(r) + V_{\textup{SD}}(r),
\label{eq2}
\end{equation}
where the reduced mass $\mu$ is defined as
$\mu = \frac{m_{\textup{cluster}}m_Q}{m_{\textup{cluster}}+m_Q}$.
Different types of potentials can be taken for $V_{\textup{SI}}(r)$, which describes the spin-independent interaction between the light quark cluster and the heavy quark. In Sec.~\ref{subsecB}, we will list four types of expression of $V_{\textup{SI}}(r)$.
By solving the Schr\"{o}dinger equation with the concrete $V_{\textup{SI}}(r)$, the excited energy $E_{nL}$ can be determined. The spin-dependent interactions $V_{\textup{SD}}(r)$ in Eq. (\ref{eq2}) include the hyperfine interaction, the spin-orbit forces, and the tensor
force \cite{Chen:2016iyi,Karliner:2017kfm}. Usually, the spin-dependent interactions are much weaker than the spin-independent interaction. Here, we may take $P$-wave charm baryon states $\Lambda_c(2595)^+$ and $\Lambda_c(2625)^+$ as an example to show it.
The mass splitting of $\Lambda_c(2595)^+$ and $\Lambda_c(2625)^+$ is about 36 MeV, which is one order smaller than their excited energies. So, the spin-dependent interactions are usually treated as the leading-order perturbation contribution in the practical calculation.

Thus, we first ignore the spin-dependent interactions and ascribe the mass gaps between the single baryons with the different strangeness but with same $nL(J^P)$ quantum numbers to the mass difference of light quark clusters.
In the qusi-two-body picture, the spin average mass of a $nL$ heavy baryon multiplet could be denoted as follows
\begin{equation}
\bar{M}_{nL} = m_Q + m_{\textup{cluster}} + E_{nL}. \label{eq3}
\end{equation}
Since two light quarks in the cluster are in the ground state, we take the chromomagnetic model~\cite{DeRujula:1975qlm,Jaffe:1976ih,Maiani:2004vq,Hogaasen:2005jv} to parameterze the mass of light quark cluster, where the mass of the light quark cluster could be written as
\begin{equation}
m_{\textup{cluster}} = m_{1} + m_{2} + A\frac{{\textbf{s}_1}\cdot{\textbf{s}_2}}{m_1m_2} \label{eq4}
\end{equation}
with
\begin{equation}
A = \frac{16\pi}{9}\langle\alpha_s(r)\delta^3(\textbf{r})\rangle. \label{eq5}
\end{equation}
In the following analysis, assuming the coefficient $A$ to be a positive value for simplicity and combing with Eqs.~(\ref{eq3}) and (\ref{eq4}), we have
\begin{equation}
\begin{aligned}\label{eq6}
\bar{M}^{\Xi_Q}_{nL}-\bar{M}^{\Lambda_Q}_{nL}=~&m_s-m_q+\frac{3A}{4m_q}\left(\frac{1}{m_q}-\frac{1}{m_s}\right)=\delta{m}+3\Delta,\\
\bar{M}^{\Xi_Q^\prime}_{nL}-\bar{M}^{\Sigma_Q}_{nL}=~&m_s-m_q-\frac{A}{4m_q}\left(\frac{1}{m_q}-\frac{1}{m_s}\right)=\delta{m}-\Delta,\\
\bar{M}^{\Omega_Q}_{nL}-\bar{M}^{\Xi_Q^\prime}_{nL}=~&m_s-m_q-\frac{A}{4m_s}\left(\frac{1}{m_q}-\frac{1}{m_s}\right)=\delta{m}-\Delta^\prime.
\end{aligned}
\end{equation}
Here, $m_q$ and $m_s$ denote the constituent masses of up/down and strange quarks, respectively, while $\delta{m}$ denotes their mass difference. Under the situations $m_s>m_q$ and $A>0$, $\delta{m}$, $\Delta$, and $\Delta^\prime$ defined in Eqs.~(\ref{eq6}) must be positive. Thus, we obtain the following relation
\begin{equation}
\bar{M}^{\Xi_Q}_{nL}-\bar{M}^{\Lambda_Q}_{nL} > \bar{M}^{\Xi_Q^\prime}_{nL}-\bar{M}^{\Sigma_Q}_{nL}, \label{eq7}
\end{equation}
which explain why the mass gaps of the baryons in the $\bar{3}_f$ multiplet
is larger than the $6_f$ multiplet well (see the comparison of the mass gap values shown in Table \ref{table1} and Table \ref{table3}). There, the measured mass gaps of the involved $\Lambda_Q$ and $\Xi_Q$ states are about 190 MeV (see Table \ref{table1}), while the mass gaps of $\Sigma_Q$ and $\Xi_Q^\prime$ states are around 120 MeV (see Table \ref{table3}).

Besides, we can do a further numerical analysis to illustrate why mass gaps of the discussed $\Omega_Q$ and $\Xi_Q^\prime$ states are around 120 MeV.
One takes the average values of mass gaps of $\Lambda_Q$ and $\Xi_Q$ baryon states with the same $nL$ to be 195 MeV, and the average values of the mass gaps of $\Sigma_Q$ and $\Xi_Q^\prime$ baryons with the same $nL$ to be 122 MeV, by which
we may fix the values of $\delta{m}$ and $\Delta$ in Eqs.~(\ref{eq6}), i.e., $\delta{m}=140.3$ MeV and $\Delta=18.3$ MeV.
If further setting the masses of light quarks as $m_q=280$ MeV and $m_s=420$ MeV, the parameters $A$ and $\Delta^\prime$ can be fixed as 1.717$\times10^7$ MeV$^3$ and 12.2 MeV, respectively.
Finally, the mass gap between $\Xi_Q^\prime$ and $\Omega_Q$ states is estimated to be
\begin{equation}
\bar{M}^{\Omega_Q}_{nL}-\bar{M}^{\Xi_Q^\prime}_{nL} \simeq 128 ~\textup{MeV}.  \label{eq8}
\end{equation}
This value is consistent with the measured mass gaps of $\Omega_Q$ and $\Xi_Q^\prime$ baryons with the same $nL(J^P)$ quantum numbers, as shown in Table \ref{table3}.

In a word, the universal behavior of mass gaps of single heavy baryons can be well understood by the constituent quark model.

\subsection{Further study of the equal mass splitting phenomenon involved in the excited $\bar{3}_f$ baryons}\label{subsecB}

Our study already indicates that the nearly equal mass gap of $\Lambda_c$ and $\Xi_c$ baryons with the same $nL$ can be reproduced.
In this subsection, we discuss the case when the spin-dependent interaction is included for the $\bar{3}_f$ baryons.

Firstly, we present the experimental results in Table \ref{table4}, where the $\Lambda_c(2595)/\Xi_c(2790)$ with $J^P=1/2^-$
and the $\Lambda_c(2625)/\Xi_c(2815)$ with $J^P=3/2^-$ are the $1P$ states.
We may find that the mass splitting of $\Lambda_c(2595)$ and $\Lambda_c(2625)$
is comparable with the splitting of $\Xi_c(2790)$ and $\Xi_c(2815)$. Similar phenomenon happens for $D$-wave $\Lambda_c$ and $\Xi_c$ baryons which are listed in the second line of Table \ref{table4}. This nearly equal mass splitting phenomenon can also be understood in our scheme.

\begin{table}[htbp]
\caption{The measured mass splittings of the 1$P$ and 1$D$ states of $\Lambda_c$ and $\Xi_c$ baryons due to the spin-orbit interaction. The corresponding spin-parity quantum numbers of these states can be found in Table \ref{table1}. }\label{table4}
\renewcommand\arraystretch{1.2}
\begin{tabular*}{86mm}{c@{\extracolsep{\fill}}ccc}
\toprule[1pt]\toprule[1pt]
 $\Lambda_Q$ states                     & splitting             & $\Xi_Q$ states    & splitting      \\
\toprule[1pt]
 $\Lambda_c(2595)^+/\Lambda_c(2625)^+$  & 35.8                  & $\Xi_c(2790)^+/\Xi_c(2815)^+$  & 24.3            \\
 $\Lambda_c(2860)^+/\Lambda_c(2880)^+$  & 25.5                  & $\Xi_c(3055)^+/\Xi_c(3080)^+$  & 21.3            \\
\bottomrule[1pt]\bottomrule[1pt]
\end{tabular*}
\end{table}

For these $\Lambda_Q$ and $\Xi_Q$ baryons in $\bar 3_f$ representation, the expression of the spin-dependent interaction is very simple since the involved spin of light quark cluster is zero. Thus, the only spin-dependent interaction is relevant to the following spin-orbit coupling
\begin{equation}
V_{\textrm{so}} = \frac{4}{3}\frac{\alpha_s}{r^3}\frac{1}{m_\textup{cluster}m_Q}\textbf{S}_Q\cdot{\textbf{L}}. \label{eq9}
\end{equation}
Here, the second and higher order contributions of $1/m_Q$ are ignored since the heavy quark mass in the baryon system is much heavier than the mass of the light quark cluster.  The spin-orbital interaction which is given in Eq.~(\ref{eq9}) has been adopted in our previous work~\cite{Chen:2014nyo} and successfully predicted the mass splitting of $D$-wave $\Lambda_c^+$ baryons.

In fact, the equal mass splitting phenomenon shown in Table \ref{table4} requires the same value of $\xi_Q=\langle\frac{4}{3}\frac{\alpha_s}{r^3}\frac{1}{m_\textup{cluster}m_Q}\rangle$ in Eq.~(\ref{eq9}) for the excited $\Lambda_Q$ and $\Xi_Q$ baryons. In the following, we will prove it.

Here, we adopt the scaling technique~\cite{Quigg:1979vr} to deal with the Schr\"{o}dinger equation, which can help us to find how the eigenvalues $E_{nL}$ and the expectation value $\langle{r^n}\rangle$ vary with changing the parameters in the quark potential model.

The Schr\"{o}dinger equation for the single heavy flavor baryon could be written as
\begin{equation}
\left[-\frac{\nabla^2}{2\mu} +V_{\textup{SI}}(r) \right]\psi^m_{nL} = E_{nL}\psi^m_{nL}.\label{eq10}
\end{equation}
As mentioned before, different kinds of potentials can be taken to depict the effective interaction between the light quark cluster and the heavy quark. We will show that the relationship $\xi_{\Lambda_Q}\simeq\xi_{\Xi_Q}$ may always stand for the usual effective potentials. To illustrate this point, we employ four typical kinds of potentials for the effective interaction $V_{\textup{SI}}(r)$, i.e., the power-law potential~\cite{Martin:1980jx}, the Cornell potential~\cite{Eichten:1978tg}, the logarithmic potential~\cite{Quigg:1977dd,Jena:1983jb}, and the Indiana potential~\cite{Fogleman:1979mr}. Their expresssions are
\begin{equation}
\begin{split}
V_1(r) = & b_1r^\nu-V_{10},\\
V_2(r) = & -\frac{4}{3}\frac{\alpha}{r}+b_2r-V_{20},\\
V_3(r) = & b_3\ln\left(\frac{r+t}{r_0}\right),\\
V_4(r) = & b_4\frac{(1-\Lambda r)^2}{r\ln r}. \label{eq11}
\end{split}
\end{equation}
More details of these phenomenological potentials could be found in Refs. \cite{Quigg:1979vr,Lucha:1991vn}.

We should repeat the nearly equal excited energies of $\Lambda_Q$ and $\Xi_Q$ states, i.e.,
\begin{equation}
E_{nL}^{\Lambda_Q} \simeq E_{nL}^{\Xi_Q}, \label{eq12}
\end{equation}
which has been implied in the discussion in Sec. \ref{subsecA} (refer to Eq.~(\ref{eq3})).
In the following, we take the power-law potential as an example to show how to obtain the following relationship
\begin{equation}
\frac{\xi_{\Lambda_Q}}{\xi_{\Xi_Q}} = \left(\frac{\mu_{\Lambda_Q}}{\mu_{\Xi_Q}}\right)^{3/2}\frac{m_{[qs]}}{m_{[ud]}}, \label{eq13}
\end{equation}
associated with the relation in Eq.~(\ref{eq12}).
Here, the $m_{[ud]}$ and $m_{[qs]}$ denote the masses of light scalar quark clusters in the $\Lambda_Q$ and $\Xi_Q$ baryons, respectively.

With the power-law potential, the radical part of Eq.~(\ref{eq10}) could be written as
\begin{equation}
-\frac{\textrm{d}^2\chi_{nl}}{\textrm{d}r^2} + \left[2\mu b_1r^\nu+\frac{l(l+1)}{r^2}\right]\chi_{nl} = 2\mu(E_{nL}+V_{10})\chi_{nl}. \label{eq14}
\end{equation}
Next, we set $z=\eta r$ and define $u_{nl}(z)\equiv\chi_{nl}(r)$. Then Eq.~(\ref{eq14}) can be translated into the following form
\begin{equation}
-\frac{\textrm{d}^2u_{nl}}{\textrm{d}z^2} + \left[\frac{2\mu b_1}{\eta^{\nu+2}}z^\nu+\frac{l(l+1)}{z^2}\right]u_{nl} = \frac{2\mu(E_{nL}+V_{10})}{\eta^2}u_{nl}.\label{eq15}
\end{equation}
When setting
\begin{equation}
\frac{2\mu b_1}{\eta^{\nu+2}} =1,~~~~~\varepsilon_{nL} = \frac{2\mu(E_{nL}+V_{10})}{\eta^2}, \label{eq16}
\end{equation}
the following scaled Schr\"{o}dinger equation is obtained
\begin{equation}
-\frac{\textrm{d}^2u_{nl}}{\textrm{d}z^2} + \left[z^\nu+\frac{l(l+1)}{z^2}\right]u_{nl} = \varepsilon_{nL}u_{nl}, \label{eq17}
\end{equation}
which can be solved by the numerical calculation or the approximation method when the parameter $\nu$ is given. For the single heavy flavor baryon, the parameter $\nu$ is constrained to be about 0.38 by the values of $\mathcal{R}_1$ and  $\mathcal{R}_2$ listed in Table \ref{table2}. The other parameters $m_{\rm cluster}$, $m_Q$, $b_1$, and  $V_{10}$ can be further constrained by fitting the masses of these known $\Lambda_Q$ and $\Xi_Q$ baryons.

In the following discussion, however, we do not need to perform the detailed numerical calculation. As pointed above, the parameter $\nu$ could be regarded as a constant for both $\Lambda_Q$ and $\Xi_Q$ states. Then the eigenvalue $\varepsilon_{nL}$ and the expectation value $\langle{z^{-3}}\rangle$ given by Eq.~(\ref{eq17}) are same for the $\Lambda_Q$ and $\Xi_Q$ states. With Eq. (\ref{eq12}), we get
\begin{equation}
A_1 = \frac{E_{nL}-E_{n^\prime L^\prime}}{\varepsilon_{nL}-\varepsilon_{n^\prime L^\prime}}=b_1(2b_1\mu)^{-\frac{\nu}{\nu+2}}, \label{eq18}
\end{equation}
which can also be regarded as a constant for the $\Lambda_Q$ and $\Xi_Q$ states. Here, Eq.~(\ref{eq18}) can be derived with the help of Eq.~(\ref{eq16}). Then one can obtain the following expression
\begin{equation}
\frac{\xi_{\Lambda_Q}}{\xi_{\Xi_Q}} = \frac{\langle r^{-3}\rangle_{\Lambda_Q}}{\langle r^{-3}\rangle_{\Xi_Q}}\frac{m_{[qs]}}{m_{[ud]}} =  \left(\frac{b_{1\Lambda_Q}\mu_{\Lambda_Q}}{b_{1\Xi_Q}\mu_{\Xi_Q}}\right)^{\frac{3}{\nu+2}}\frac{m_{[qs]}}{m_{[ud]}} \label{eq19}
\end{equation}
with help of $\langle r^{-3}\rangle=\eta^3\langle z^{-3}\rangle=(2b\mu)^{3/(\nu+2)}\langle z^{-3}\rangle$. In addition, one deduces $b_1$ by Eq.~(\ref{eq18}) as
\begin{equation}
b_1 = A^{\frac{\nu+2}{2}}(2\mu)^{\nu/2}. \label{eq20}
\end{equation}
Combing Eq.~(\ref{eq19}) and Eq. (\ref{eq20}), the result of Eq.~(\ref{eq13}) can be obtained. The similar derivations involved in other three kinds of potentials are presented in Appendix. With the definition of $\mu$, we further have the following relations
\begin{equation}
\frac{\xi_{\Lambda_Q}}{\xi_{\Xi_Q}} = \left(\frac{m_Q+m_{[qs]}}{m_Q+m_{[ud]}}\right)^{3/2}\left(\frac{m_{[ud]}}{m_{[qs]}}\right)^{1/2}.\label{eq21}
\end{equation}
When taking the following values
\begin{equation}
\begin{split}
m_c=1.55~\textrm{GeV},~~~~~~~~&m_{[ud]}=0.71~\textrm{GeV},\\
m_b=4.65~\textrm{GeV},~~~~~~~~&m_{[us]}=0.90~\textrm{GeV}, \label{eq22}
\end{split}
\end{equation}
for the masses of light quark clusters and heavy quarks in the single heavy flavor baryon, we estimate the values of two ratios
\begin{equation}
\frac{\xi_{\Lambda_c}}{\xi_{\Xi_c}} \simeq 1.003,~~~~~~~\frac{\xi_{\Lambda_b}}{\xi_{\Xi_b}} \simeq 0.936,  \label{eq23}
\end{equation}
which prove that $\xi_Q$ almost keeps same for the $\Lambda_Q$ and $\Xi_Q$ baryons.

\section{Application}\label{sec3}

The universal behavior of mass gaps which was discussed above can give some valuable implications, especially, for predicting the masses of the undiscovered single heavy baryon states.

 \begin{itemize}
 \item[(1)]

A neutral resonance was discovered by the BaBar Collaboration in the process $B^-\to\Sigma_c(2846)^0\bar{p}\to\Lambda_c^+\pi^-\bar{p}$~\cite{Aubert:2008ax}. Its mass and decay width are
\begin{equation}
\begin{aligned}\label{eq24}
m(\Sigma_c(2846)^0) ~=~& 2846\pm8\pm10 ~\textrm{MeV},\\
\Gamma(\Sigma_c(2846)^0) ~=~& 86^{+33}_{-22}\pm12 ~\textrm{MeV}.
\end{aligned}
\end{equation}
With the higher mass and the weak evidence of $J = 1/2$, BaBar suggested the signal of $\Sigma_c(2846)^0$ to be different
from the $\Sigma_c(2800)^{0,+,++}$ states which were discovered by the Belle Collaboration via the $e^+e^-$  collision~\cite{Mizuk:2004yu}. For the limited  information of the $J^P$ quantum numbers and branching ratios, however, PDG temporarily treated the $\Sigma_c(2846)^0$ and $\Sigma_c(2800)^{0,+,++}$ as a same state~\cite{Zyla:2020zbs}.

The universal behavior of mass gap discussed in this work may provide some valuable clues for clarifying the puzzle of $\Sigma_c(2846)^0$ and $\Sigma_c(2800)^{0,+,++}$ states. We find that the recently reported $\Xi_c^\prime(2965)^0$~\cite{Aaij:2020yyt} and $\Omega_c(3090)^0$~\cite{Aaij:2017nav,Yelton:2017qxg} could form the following chain
\begin{equation}
\Sigma_c(2846)^0~\leftrightarrow~~\Xi_c^\prime(2965)^0~\leftrightarrow~~\Omega_c(3090)^0, \label{eq25}
\end{equation}
if adding $\Sigma_c(2846)^0$, which satisfies the requirement from universal behavior of mass gaps.
Additionally, the $\Sigma_c(2800)^{0,+,++}$ has been grouped into another chain with $\Xi_c^\prime(2923)^0$ and $\Omega_c(3050)^0$, as shown in Table \ref{table3}. This analysis based on universal behavior of mass gap suggests that the $\Sigma_c(2846)^0$ and $\Sigma_c(2800)^{0,+,++}$ baryons should be two different states.

 \item[(2)]
As the second example, we will apply the universal behavior of mass gap rule to predict the masses of 1$P$, $1D$, and 2$S$ $\Xi_b$ states. At present, only the ground states, i.e., the $\Xi_b(5792)^0$ and the $\Xi_b(5797)^-$, have been established. Since LHCb has shown its capability in accumulating the data sample of the excited bottom baryon resonances in the past years, we expect the following predicted masses of excited $\Xi_b^0$ baryons to be tested by the LHCb Collaboration in the near future.

The mass gap between $\Xi_b(5792)^0$ and $\Lambda_b(5620)^0$ state is 172.3 MeV which is about 10 MeV smaller than the mass difference of $\Lambda_c(2286)^+$ and $\Xi_c(2470)^+$ states. Comparing the results in Table \ref{table1}, we may take the mass gaps as 185 MeV for the excited $\Lambda_b^0$ and $\Xi_b^0$. With the measured masses of $\Lambda_b(6072)^0$ states, the mass of 2$S$ $\Xi_b$ state could be predicted as 6257 MeV directly. We have shown that the mass splitting of $nL$ $\Xi_Q$ states is nearly equal to the corresponding $\Lambda_Q$ baryons (see Eq.~(\ref{eq23})). Therefore, the masses of two 1$P$ $\Xi_b$ states are predicted to be 6097 MeV and 6105 MeV, while the masses of two 1$D$ $\Xi_b$ states are about 6331 MeV and 6337 MeV.

\item[(3)]
As the last example, we point out that two resonance structures may exist in the previously observed signals of $\Sigma_b(6097)^\pm$~\cite{Aaij:2018tnn} and $\Xi_b^\prime(6227)^{0,-}$~\cite{Aaij:2018yqz,Aaij:2020fxj}. Two narrow bottom baryons $\Omega_b(6340)^-$ and $\Omega_b(6350)^-$ which were reported by LHCb in the $\Xi_b^0K^-$ decay channel~\cite{Aaij:2020cex} could be regarded as the good candidates of $P$-wave states with $J^P=3/2^-$ and $J^P=5/2^-$, respectively. With this assignment, the mass splitting of these two states can also be explained by the quark potential model~\cite{Ebert:2011kk,Chen:2018vuc} and the QCD sum rule~\cite{Yang:2020zrh}. According to the universal behavior of mass gaps of single heavy flavor baryons, we could conjecture that the $\Sigma_b$ and $\Xi_b^\prime$ partners of $\Omega_b(6340)^-$ and $\Omega_b(6350)^-$ have been contained in the signals of $\Sigma_b(6097)^\pm$ and $\Xi_b^\prime(6227)^{0,-}$. The situation of charmed baryons is alike. The observed $\Xi_c(2923)^0$ and $\Omega_c(3050)^0$ can be explained as the $P$-wave states with $J^P=3/2^-$, while the $\Xi_c(2939)^0$ and $\Omega_c(3065)^0$ could be regarded as the $J^P=5/2^-$ partners. Thus, we may point out that the signal of $\Sigma_c(2800)^{0,+,++}$ states~\cite{Mizuk:2004yu} may also contain the $J^P=3/2^-$ and $J^P=5/2^-$ states. With the higher statistical precision, we expect the LHCb and Belle II experiments to distinguish these two resonance structures in future.

\end{itemize}

\section{Further discussion for the $\rho$-mode excited single heavy baryons}\label{sec4}

It is interesting to point out that none of $\rho$-mode excited single heavy baryons have been established by experiments. We have treated these discovered heavy baryons as the $\lambda$-mode excitations and explained their mass gaps well. One may naturally want to know how about the mass gaps of these $\rho$-mode excited single heavy baryons. As shown in Sec.~\ref{sec3}, the mass gaps of these $\lambda$-mode excited heavy baryons arise from the mass difference of $s$ and $u/d$ quarks and the chromomagnetic interaction of two light quarks in the cluster. For the $\rho$-mode excited single heavy baryons, the chromomagnetic interaction of two light quarks becomes smaller and can be negligible. Then the mass gap of corresponding $\rho$-mode excited single heavy baryons mainly comes from the mass difference of $s$ and $u/d$ quarks. This means that the mass gap behavior of $\rho$-mode excited single heavy baryons should be different from these $\lambda$-mode excited heavy baryons.

For confirming our conjecture above, we will calculate the masses of charm baryons by a non-relativistic quark model in the three-body picture. The Hamiltonian of this model is given as~\cite{Luo:2021dvj}
\begin{equation}\label{eq26}
\hat{H}_0=\sum\limits_{i=1}^{3}\left(m_i+\frac{p_i^2}{2m_i}\right)+\sum\limits_{i<j}V_{ij},
\end{equation}
where $m_i$ and $p_i$ are the mass and momentum of $i$-th constituent quark. The $V_{ij}$ in Eq.~(\ref{eq26}) denotes the interactions between quark-quark in the baryon system, which contains both spin-independent and spin-dependent interactions. Specifically, the $V_{ij}$ is written as
\begin{equation}\label{eq27}
V_{ij}=-\frac{2}{3}\frac{\alpha_s}{r_{ij}}+\frac{1}{2}br_{ij}-C+\frac{16\pi\alpha_s}{9m_im_j}\left(\frac{\sigma}{\sqrt{\pi}}\right)^3{\rm e}^{-\sigma^2r^2}{\bf s}_i\cdot {\bf s}_j+V^{\rm tens}_{ij}+V^{\rm so}_{ij},
\end{equation}
where the $\alpha_s$, $b$, $\sigma$, and $C$ denote the coupling constant of one-gluon exchange (OGE), the strength of linear confinement, the Gaussian smearing parameter, and a mass-renormalized constant, respectively. The $V^{\rm tens}_{ij}$ and $V^{\rm so}_{ij}$ in Eq.~(\ref{eq27}) are the tensor and spin-orbit terms, respectively (see Ref.~\cite{Capstick:1986bm} for more details). The model parameters are collected in Table~\ref{table5}.
\begin{table}[htbp]
\centering
\caption{The parameters of non-relativistic quark potential model. The consistent quark masses are presented in the last row.}
\label{table5}
\renewcommand\arraystretch{1.25}
\begin{tabular*}{86mm}{@{\extracolsep{\fill}}
lcccc
}
\toprule[1.00pt]
\toprule[1.00pt]
                     &$\alpha_s$ &$b$ (GeV$^2$) &$C$ (GeV) &$\sigma$ (GeV)  \\
\midrule[0.75pt]
$\Lambda_c/\Sigma_c$ &0.580      &0.117         & 0.219    &1.220           \\
$\Xi_c/\Xi_c^\prime$ &0.580      &0.132         & 0.220    &1.220           \\
\midrule[0.75pt]
\multicolumn{5}{c}{$m_{u/d}=0.310$ GeV~~~$m_s=0.450$ GeV~~~$m_c=1.650$ GeV} \\
\bottomrule[1.00pt]
\bottomrule[1.00pt]
\end{tabular*}
\end{table}

As shown in Table~\ref{table6}, the measured masses of well-established charm baryons can be reproduced by the non-relativistic quark potential model.

\begin{table}[htbp]
\caption{A comparison of the predicted masses of established charm baryons with the measured results (in MeV). The measured masses are listed in the first row below the states while the predictions are given in the second row. The measured masses of $\Xi^\prime_c(2923)$, $\Xi^\prime_c(2939)$, and $\Xi^\prime_c(2965)$ are taken from Ref.~\cite{Aaij:2020yyt}, while the others are borrowed from Ref.~\cite{Zyla:2020zbs}.}\label{table6}
\renewcommand\arraystretch{1.2}
\begin{tabular*}{85mm}{c@{\extracolsep{\fill}}ccccc}
\toprule[1pt]\toprule[1pt]
 $\Lambda_c(2286)$          & $\Lambda_c(2765)$     & $\Lambda_c(2595)$  & $\Lambda_c(2625)$  & $\Lambda_c(2860)$   & $\Lambda_c(2880)$ \\
\toprule[0.8pt]
      2286.5                & 2766.6                & 2592.3             & 2628.1             & 2856.1              & 2881.6       \\
      2286                  & 2788                  & 2595               & 2620               & 2858                & 2871       \\
\toprule[0.8pt]
 $\Xi_c(2468)$              & $\Xi_c(2970)$         & $\Xi_c(2790)$      & $\Xi_c(2815)$      & $\Xi_c(3055)$       & $\Xi_c(3080)$ \\
\toprule[0.8pt]
      2467.9                & 2966.3                & 2792.4             & 2816.7             & 3055.9              & 3077.2       \\
      2466                  & 2985                  & 2786               & 2811               & 3060                & 3071       \\
\toprule[0.8pt]
 $\Sigma_c(2455)$           & $\Sigma_c(2520)$      & $\Sigma_c(2800)$   &                    &                     &                   \\
\toprule[0.8pt]
      2454.0                & 2518.4                & 2801.0             &                    &                     &       \\
      2463                  & 2511                  & 2791               &                    &                     &          \\
\toprule[0.8pt]
 $\Xi^\prime_c(2580)$       & $\Xi^\prime_c(2645)$  & $\Xi^\prime_c(2923)$ & $\Xi^\prime_c(2939)$ & $\Xi^\prime_c(2965)$ &   \\
\toprule[0.8pt]
      2578.4                & 2645.6                & 2923.0             & 2938.6             & 2964.9              &        \\
      2595                  & 2648                  & 2928               & 2949               & 2934                &        \\
\bottomrule[1pt]\bottomrule[1pt]
\end{tabular*}
\end{table}

The non-relativistic quark model also allows us to obtain the masses of $\rho$-mode excited charm baryons simultaneously with the same parameters in Table~\ref{table5}. For illustrating the mass gaps of $\rho$-mode excited charmed and charmed-strange baryons, we list the predicted masses and mass gaps in Table~\ref{table7}.
\begin{table}[htbp]
\caption{The predicted masses of $P$-wave $\rho$-mode excited charm baryons (in MeV). States with the same $J^P$ are distinguished by their different masses. Specifically, the states with the lower and higher masses are denoted by the subscripts ``$L$'' and ``$H$'', respectively.}\label{table7}
\renewcommand\arraystretch{1.2}
\begin{tabular*}{85mm}{c@{\extracolsep{\fill}}ccccc}
\toprule[1pt]\toprule[1pt]
  & $|1/2^-\rangle_L$ & $|1/2^-\rangle_H$ &$|3/2^-\rangle_L$  & $|3/2^-\rangle_H$ & $|5/2^-\rangle$  \\
\toprule[0.8pt]
  $\Lambda^\rho_c(1P)$    &  2862                & 2868                & 2834             & 2891             & 2863                    \\
  $\Xi^\rho_c(1P)$        &  3010                & 3016                & 2988             & 3048             & 3021                      \\
  $\Delta{M}$             &  148                 & 148                 & 154              & 157              & 158                      \\
\toprule[0.8pt]
  & $|1/2^-\rangle$ & $|3/2^-\rangle$ &   &   &    \\
\toprule[0.8pt]
  $\Sigma^\rho_c(1P)$     &   2854               & 2874                &               &                 &                     \\
  $\Xi^{\prime\rho}_c(1P)$ &  3005               & 3027                &               &                 &                       \\
  $\Delta{M}$              &  151                & 153                 &               &                 &                       \\
\bottomrule[1pt]\bottomrule[1pt]
\end{tabular*}
\end{table}

For the $P$-wave $\rho$-mode excited charmed baryons, there are five $\Lambda_c$ states and two $\Sigma_c$ states (see Table~\ref{table7}). The case of charmed-strange baryons is alike. By comparing of the predicted masses, one may find that the mass gaps of $P$-wave $\rho$-mode excited charmed and charmed-strange baryons are all around 150 MeV. As shown in Table~\ref{table5}, the mass of $s$ quark in our model is 140 MeV larger than the $u/d$ quark. Then we may preliminarily conclude that the mass gaps of $\rho$-mode excited charm baryons mainly come from the mass difference of $s$ and $u/d$ quarks. Therefore, we confirmed our conjecture above.

\section{Summary}\label{sec5}

Until now, about fifty single heavy baryons have been observed by experiment \cite{Zyla:2020zbs}, where their isospin partners have not been counted. Facing such abundant experimental data of single heavy flavor baryons, we have a good chance to check the mass gaps existing in the established mass spectrum of single heavy flavor baryons.

In this work, we analysed the measured masses of these discovered single heavy baryons and pointed out that some universal behaviors of the mass gaps may exist in this kind of baryons. By a constituent quark model,
we have revealed the underlying mechanism behind this universal phenomenon and explained the universal behaviors of mass gaps for the $\bar 3_f$ and $6_f$ single heavy flavor baryons. Additionally, we also illustrate why there exists the nearly equal mass splitting for the orbital excited baryons in the $\bar 3_f$ representation. It is important to note that these discovered heavy baryons are assigned as the $\lambda$-mode excitations.

Universal behaviors of mass gaps can be applied to the mass spectrum analysis. In this work, we
give three implications: 1) We indicated that the neutral resonance $\Sigma_c(2846)^0$ reported from the BaBar Collaboration \cite{Aubert:2008ax} is different from the $\Sigma_c(2800)^{0,+,++}$ states \cite{Mizuk:2004yu}. 2) The masses of 1$P$, 1$D$, and 2$S$ $\Xi_b^0$ states were predicted. 3) We pointed out that two resonance structures may exist in the previously observed signals of $\Sigma_c(2800)^{0,+,++}$~\cite{Zyla:2020zbs}, $\Sigma_b(6097)^\pm$ \cite{Aaij:2018tnn}, and $\Xi_b^\prime(6227)^{0,-}$ \cite{Aaij:2018yqz,Aaij:2020fxj} states.
These predictions based on the universal behaviors of mass gaps could be tested by the LHCb and Belle II experiments in future.

We also discuss the mass gaps of these $\rho$-mode excited single heavy flavor baryons. We find that the mass gaps of $P$-wave $\rho$-mode excited charmed baryons are around 150 MeV. Then we may preliminarily conclude that the mass gap behavior of $\rho$-mode excited single heavy baryons is quite different from these $\lambda$-mode excitations. So the mass gaps may provide some valuable clues for distinguishing the $\rho$-mode excited single heavy baryons from the $\lambda$-mode excitations in future.

\section*{ACKNOWLEDGMENTS}

X. L. is supported by the China National Funds for Distinguished Young Scientists under Grant No. 11825503, National Key Research and Development Program of China under Contract No. 2020YFA0406400 and the 111 Project under Grant No. B20063. B. C. is partly supported by the National Natural Science Foundation of China under Grants No. 11305003, No. 11647301, and No. 12047501.
\vspace{0.3 cm}

$Note~added.-$After the submission of this paper to the arXiv, a new bottom baryon state, namely, the $\Xi_b(6100)^-$, was reported by the CMS Collaboration~\cite{Sirunyan:2021vxz}. Its mass is in good agreement with the predictions of $P$-wave $\Xi_b$ states in Sec.~\ref{sec3}. In fact, this newly observed $\Xi_b$ state has been suggested to be the $P$-wave candidate with $J^P=3/2^-$ in Ref.~\cite{Sirunyan:2021vxz}.

\section*{APPENDIX}\label{Appendix1}

In this section, we follow the similar derivations presented in Sec.~\ref{subsecB} and show how to obtain the relation in Eq.~(\ref{eq13}) by the Cornell potential, the logarithmic potential, and the Indiana potential.

For the Cornell potential, the Schr\"{o}dinger equation is
\begin{equation}
-\frac{\textrm{d}^2\chi_{nl}}{\textrm{d}r^2} + \left[2\mu\left(-\frac{4}{3}\frac{\alpha}{r}+b_2r\right)+\frac{l(l+1)}{r^2}\right]\chi_{nl} = 2\mu(E_{nL}+V_{20})\chi_{nl}. \label{A1}
\end{equation}
When we set $z=\eta r$ and define $u_{nl}(z)\equiv\chi_{nl}(r)$, Eq.~(\ref{A1}) can be translated into the following form
\begin{equation}
-\frac{\textrm{d}^2u_{nl}}{\textrm{d}z^2} + \left[\left(\frac{2\mu b_2}{\eta^3}z-\frac{8\alpha\mu}{3\eta z}\right)+\frac{l(l+1)}{z^2}\right]u_{nl} = \frac{2\mu(E_{nL}+V_{20})}{\eta^2}u_{nl}. \label{A2}
\end{equation}
We further set $\frac{8\alpha\mu}{3\eta} = 1$, $\frac{2\mu b_2}{\eta^3} = \lambda$, and $\varepsilon_{nL} = \frac{2\mu(E_{nL}+V_{20})}{\eta^2}$, the scaled Schr\"{o}dinger equation for the Cornell potential is given as
\begin{equation}
-\frac{\textrm{d}^2u_{nl}}{\textrm{d}z^2} + \left[\left(-\frac{1}{z}+\lambda z\right)+\frac{l(l+1)}{z^2}\right]u_{nl} = \varepsilon_{nL}u_{nl}. \label{A3}
\end{equation}
The only parameter $\lambda$ in Eq.~(\ref{A3}) could be fixed as $\lambda \simeq 1.70$ by the $\mathcal{R}_1$ and  $\mathcal{R}_2$ (see Eq.~(\ref{eq1}) and Table \ref{table2}) for $\Lambda_Q$ and $\Xi_Q$ baryons. With the variable substitutions above, we have
\[\begin{cases}
~~~~r      &= ~\dfrac{3}{8\alpha\mu}z,\\
~~~~b      &= ~\dfrac{2^8\mu^2\alpha^3}{27}\lambda,\\
~E_{nL}&= ~\dfrac{32\mu\alpha^2}{9}\varepsilon_{nL}-V_{20}.
\end{cases}\]
Due to the condition of Eq.~(\ref{eq12}), the following ratio
\begin{equation}
A_2 = \frac{E_{nL}-E_{n^\prime L^\prime}}{\varepsilon_{nL}-\varepsilon_{n^\prime L^\prime}} = \dfrac{32\mu\alpha^2}{9}. \label{A4}
\end{equation}
could be regarded as a constant for the $\Lambda_Q$ and $\Xi_Q$ baryons. For the Cornell potential, we have
\begin{equation}
\frac{\xi_{\Lambda_Q}}{\xi_{\Xi_Q}} = \frac{\langle r^{-3}\rangle_{\Lambda_Q}}{\langle r^{-3}\rangle_{\Xi_Q}}\frac{m_{[qs]}}{m_{[ud]}} =  \left(\frac{\mu_{\Xi_Q}\alpha_{\Xi_Q}}{\mu_{\Lambda_Q}\alpha_{\Lambda_Q}}\right)^3\frac{m_{[qs]}}{m_{[ud]}}. \label{A5}
\end{equation}
Then Eq.~(\ref{eq13}) can be obtained directly by considering the relationship $\alpha=\frac{3}{4}\sqrt{\frac{A_2}{2\mu}}$.

For the logarithmic potential, the Schr\"{o}dinger equation is given as
\begin{equation}
-\frac{\textrm{d}^2\chi_{nl}}{\textrm{d}r^2} + \left[2\mu b_3\ln\left(\frac{r+t}{r_0}\right)+\frac{l(l+1)}{r^2}\right]\chi_{nl} = 2\mu E_{nL}\chi_{nl}. \label{A6}
\end{equation}

Next, we set $z=\eta r$ and define $u_{nl}(z)\equiv\chi_{nl}(r)$. Then Eq.~(\ref{A6}) can be translated into the following form
\begin{equation}
-\frac{\textrm{d}^2u_{nl}}{\textrm{d}z^2} + \left[\ln\left(z+z_0\right)+\frac{l(l+1)}{z^2}\right]u_{nl} = \varepsilon_{nL}u_{nl}. \label{A7}
\end{equation}
where we have done the following variable substitutions
\[\begin{cases}
~~\eta    &= ~\sqrt{2b_3\mu},\\
~~z_0     &= ~\sqrt{2b_3\mu}t,\\
~E_{nL}   &= ~b_3\left[\varepsilon_{nL}-\ln\left(\sqrt{2b_3\mu}r_0\right)\right].
\end{cases}\]
For the $\Lambda_Q$ and $\Xi_Q$ baryons, the parameter $z_0$ in the scaled Schr\"{o}dinger equation of Eq.~(\ref{A7}) can be fixed as $z_0\simeq1.80$ by the $\mathcal{R}_1$ and  $\mathcal{R}_2$. Similar to the Eq.~(\ref{eq18}), one could treat the parameter $b_3$ as a constant for the $\Lambda_Q$ and $\Xi_Q$ baryons. Thus, Eq.~(\ref{eq13}) can be obtained directly.

Finally, the Schr\"{o}dinger equation of the Indiana potential is given as
\begin{equation}
-\frac{\textrm{d}^2\chi_{nl}}{\textrm{d}r^2} + \left[2\mu b_4\frac{(1-\Lambda r)^2}{r\ln(\Lambda r)}+\frac{l(l+1)}{r^2}\right]\chi_{nl} = 2\mu E_{nL}\chi_{nl}. \label{A8}
\end{equation}
In the same way, we set $z=\eta r$ and define $u_{nl}(z)\equiv\chi_{nl}(r)$. Eq.~(\ref{A8}) becomes as
\begin{equation}
-\frac{\textrm{d}^2u_{nl}}{\textrm{d}z^2} + \left[\frac{2\mu b_4}{\eta}\frac{(1-\frac{\Lambda}{\eta}z)^2}{z\ln(\frac{\Lambda}{\eta}z)}+\frac{l(l+1)}{z^2}\right]u_{nl} = \frac{2\mu E_{nL}}{\eta^2}u_{nl}. \label{A9}
\end{equation}
When one sets $\eta=\Lambda$ and does the following variable substitutions
\begin{equation}
\kappa = \frac{2\mu b_4}{\Lambda}, ~~~~~~~ E_{nL} = \frac{\Lambda^2}{2\mu}\varepsilon_{nL}, \label{A10}
\end{equation}
the scaled Schr\"{o}dinger equation for the Indiana potential is given as
\begin{equation}
-\frac{\textrm{d}^2u_{nl}}{\textrm{d}z^2} + \left[\kappa\frac{(1-z)^2}{z\ln{z}}+\frac{l(l+1)}{z^2}\right]u_{nl} = \varepsilon_{nL}u_{nl}. \label{A11}
\end{equation}

For the $\Lambda_Q$ and $\Xi_Q$ baryons, the parameter $\kappa$ in the scaled Schr\"{o}dinger equation above should be fixed as $\kappa\simeq0.80$ by the $\mathcal{R}_1$ and  $\mathcal{R}_2$. Similar to the Eq.~(\ref{eq18}), the following ratio
\begin{equation}
A_4 = \frac{E_{nL}-E_{n^\prime L^\prime}}{\varepsilon_{nL}-\varepsilon_{n^\prime L^\prime}} = \frac{\Lambda^2}{2\mu} \label{A12}
\end{equation}
could be treated as a constant for the $\Lambda_Q$ and $\Xi_Q$ baryons. With the $\eta=\Lambda=\sqrt{2A_4\mu}$, Eq.~(\ref{eq13}) could be obtained for the Indiana potential.


\end{document}